\documentclass[fleqn,usenatbib]{mnras}

\usepackage[varg]{newtxmath}
\usepackage{newtxtext}
\usepackage[T1]{fontenc}

\DeclareRobustCommand{\VAN}[3]{#2}
\let\VANthebibliography\thebibliography
\def\thebibliography{\DeclareRobustCommand{\VAN}[3]{##3}\VANthebibliography}

\usepackage{graphicx}	% Including figure files
\usepackage{amsmath}	% Advanced maths commands
\usepackage{bm, listings}
\usepackage{color}

\newcommand{\func}[2]{ #1 \left( #2 \right) }
\newcommand{\scr}[2]{ #1_{\mathrm{#2}} }

\newcommand{\diff}[2]{ \frac{\mathrm{d} #1}{\mathrm{d} #2} }

\newcommand{\partdiff}[2]{ \frac{\partial #1}{\partial #2} }

\newcommand{\ve}[1]{ \bm{\mathrm{#1}} }
\newcommand{\unitvector}[1]{\ve{\hat{e}}_{#1}}
\newcommand{\matri}[1]{\underset{\sim}{\ve{#1}}}

\newcommand{\grad}{\ve{\nabla}}
\newcommand{\divF}[1]{\grad \ve{. \: #1}}

\newcommand{\Msun}{M_{\odot}}           % Solar mass
\newcommand{\Mearth}{M_{\oplus}}        % Earth mass
\newcommand{\Mjup}{\scr{M}{Jup}}     % Jupiter mass
\newcommand{\Mstar}{\scr{M}{\ast}}   % star's mass

\newcommand{\Mpl}{\scr{M}{pl}}       % planet's mass
\newcommand{\apl}{\scr{a}{pl}}       % planet's orbital separation
\newcommand{\Rpl}{\scr{R}{pl}}       % planet's radius

\newcommand{\RHill}{\scr{R}{Hill}}   % Hill radius
\newcommand{\Lsun}{L_{\odot}}           % Solar luminosity

\newcommand{\metre}{\ \mathrm{m}}
\newcommand{\km}{\ \mathrm{km}}
\newcommand{\um}{\ \mu \mathrm{m}}
\newcommand{\millim}{\ \mathrm{mm}}
\newcommand{\kg}{\ \mathrm{kg}}
\newcommand{\kelvin}{\ \mathrm{K}}
\newcommand{\AU}{\ \mathrm{AU}}

\newcommand{\yr}{\ \mathrm{yr}}

\newcommand{\radians}{\ \mathrm{rad}}

\newcommand{\ordinal}[1]{$^{ \mathrm{#1} }$}
\newcommand{\OmegaK}{\scr{\Omega}{K}}

\graphicspath{ {Paper1Graphics/} }

\title[Size-selective accretion of dust onto CPDs]{Size-selective accretion of dust onto CPDs: Low CPD masses and filtration of larger grains}

\author[S. M. Karlin et al.]{
Samuel M. Karlin,$^{1}$\thanks{Email: s.m.karlin@gmail.com}
Olja Pani\'c,$^{1}$
and Sven van Loo$^{1,2}$
\\
$^{1}$School of Physics and Astronomy, University of Leeds, Woodhouse Lane, Leeds LS2 9JT, UK\\
$^{2}$Department of Applied Physics, Ghent University, Sint-Pietersnieuwstraat 41, Technicum blok 4
9000 Gent, Belgium \\
}

\date{Accepted 2023 January 12. Received 2022 December 22; in original form 2022 August 10}

\pubyear{2023}

\begin{document}
\label{firstpage}
\pagerange{\pageref{firstpage}--\pageref{lastpage}}
\maketitle

\begin{abstract}
The major satellites of Jupiter and Saturn are believed to have formed in circumplanetary discs, which orbit forming giant protoplanets. Gas and dust in CPDs have different distributions and affect each other by drag, which varies with grain size. Yet simulations of multiple dust grain sizes with separate dynamics have not been done before. We seek to assess how much dust of each grain size there is in circumplanetary discs. We run multifluid 3D hydrodynamical simulations including gas and four discrete grain sizes of dust from $1\um$ to $1\millim$, representing a continuous distribution. We consider a $1\Mjup$ protoplanet embedded in a protoplanetary disc around a $1\Msun$ star. Our results show a truncated MRN distribution at smaller grain sizes, which starts to tail off by $a=100\um$ and is near zero at $1\millim$. Large dust grains, which hold most of the dust mass, have very inefficient accretion to the CPD, due to dust filtration. Therefore CPDs' dust masses must be small, with mass ratio $\sim$ a few $\times 10^{-6}$ to the protoplanet. These masses and the corresponding millimetre opacities are in line with CPD fluxes observed to date.
\end{abstract}


\section{Introduction}
\label{sec:introduction}
The major satellites of Jupiter and Saturn exhibit almost perfectly coplanar prograde circular orbits, with remarkably low eccentricities and inclinations. This lends itself to the suggestion that they formed in discs of gas and dust orbiting around their parent planets \citep{Korycansky1991, WardCanup2010}. These circumplanetary discs (CPDs) are thus the birthplaces of icy moons such as Europa and Enceladus, considered promising candidates for extraterrestrial life \citep{Greenberg2011, Blanc2020, Parkinson2008, Neveu2020}. They also regulate the flow of material onto a protoplanet \citep{Rivier2012}, from which it follows that they determine the final mass that the mature planet can attain. Circumplanetary discs used to be a prediction of theorists alone, but in recent years, with VLT K-band observations of the protoplanet PDS 70 b \citep{Christiaens2019} and ALMA submillimetre observations of the protoplanet PDS 70 c \citep{Isella2019, Benisty2021}, emission from CPDs has begun to be directly observed. As such, study of CPDs is both pertinent and timely.

The dynamics of differently-sized dust particles in circumplanetary discs remains an understudied topic. The grain size of dust in CPDs is important in multiple ways. It will govern their resulting opacity, in which dust, despite being greatly outmassed by gas, is the dominant component \citep{WilliamsCieza2011}. That means it governs their temperature, which is crucial to our ability, or lack thereof, to detect CPDs observationally. Furthermore, dust size has implications for the feasibility of satellitesimal formation from dust, and thus of satellite formation.

The earliest CPD modelling, done by \citet{LunineStevenson1982}, takes the observed mass of Jupiter's Galilean satellites, multiplies it by $100$ to adjust for a dust-to-gas ratio of $10^{-2}$ and concludes that Jupiter's circumplanetary disc must have had a minimum mass of $\sim 0.02$ times the mass of Jupiter. Applying the same reasoning to the Saturnian system gives a strikingly similar fraction. The combined satellites of each planet have about $2 \times 10^{-4}$ times the mass of the planet \citep{CanupWard2009}. \citet{Mosqueira2003} point out that a disc as dense as this `rich disc' model proposes would drag satellitesimals into the protoplanet at extremely short migration timescales ($< 10^3 \yr$), rendering the formation of satellites like those of Saturn and Jupiter unfeasible. They suggest an alternative `gas-starved disc' model, where the CPD is continuously being fed more matter from outside and losing matter to the protoplanet. The observations of \citet{Isella2019} and \citet{Benisty2021} constrain the dust masses of circumplanetary discs around protoplanets in the PDS 70 system. With the caveat that this is only one star-system and it may not be representative, their results tentatively suggest that observed dust masses seem too low for the rich disc models. The starved disc models are a better fit.

Circumplanetary discs form inside gaps in the parent protoplanetary disc; a gap is a necessary but not sufficient condition for a CPD to exist. Only a sufficiently massive protoplanet can exert sufficiently strong gravitational torque to form a gap;
\begin{equation}
\Mpl > 0.39\Mjup \times \frac{\Mstar}{\Msun} \times \left(\frac{H/R}{0.05}\right)^3
\label{eq:gap_mpl} \end{equation}
\citep{LinPapaloizou1993} where $\Mpl$ and $\Mstar$ are the masses of the protoplanet and star, $\Mjup$ and $\Msun$ are the masses of Jupiter and the Sun, and $H/R$ is the protoplanetary disc's aspect ratio at the protoplanet's location. Earth-like protoplanets are thus precluded from having circumplanetary discs, but giant protoplanets ought to.

Several authors have perfomed three-dimensional single-fluid hydrodynamical simulations of CPDs: \citet{Bate2003}, \citet{Machida2008}, \citet{Tanigawa2012}, \citet{Szulagyi2014}, \citet{SzulagyiMasset2016}, \citet{SzulagyiMordasini2017}, \citet{Szulagyi2017} to name but a few. Among their many results they find that most of the protoplanet's mass inflow comes vertically from above and below the midplane, not through the CPD on the midplane. This renders two-dimensional simulations impractical to capture protoplanet growth. CO gas velocity observations by \citet{Teague2019} affirm simulations' prediction that protoplanets should have these meridional flows. Another conclusion from these simulations is that viscosity has a strong effect on the resultant accretion rate: more viscous CPDs grant their protoplanets faster accretion. The artificial numerical viscosity of simulations can be a problem for modelling low-viscosity cases \citep{Szulagyi2014} because it means that simulations intended to be inviscid, or nearly so, might be quite viscous in practice. The temperature of the CPD and of the protoplanet also plays a major role in determining the shape of the CPD and gap, the mass of the CPD, and even whether or not a CPD can exist at all. A hot enough protoplanet will have no CPD, only a gaseous envelope filling the whole Roche lobe \citep{SzulagyiMasset2016}.

However, in the literature, circumplanetary discs have been simulated assuming that gas and dust have the same distribution in space. The assumption is that they have the same dust-to-gas ratio, typically the interstellar medium value of $10^{-2}$ \citep{Knapp1974}, at all points in space. Sometimes this assumption is not made explicit; it is implicit in the work by using opacity tables which assume a dust-to-gas ratio of $10^{-2}$. It is well-known observationally that gas and dust do not share the same distribution in space in protoplanetary discs (e.g. \citealt{Long2018}; \citealt{Pinte2016}; among many others). According to theory they do not obey the same physics, so there is no reason to expect them to.

The separate dynamics of dust from gas should not be neglected. Even if it is only 1\% of the mass budget as per the ISM dust-to-gas ratio, the dust plays an outsized role in heating and cooling because it dominates the opacity: $\left(\kappa\rho\right)_d \gg \left(\kappa\rho\right)_g$ despite $\rho_g \gg \rho_d$ \citep{WilliamsCieza2011} where $\kappa$ is opacity and $\rho$ is density and $g$ and $d$ denote gas and dust. For the same reason of high opacity, dust emits disproportionately much of the electromagnetic radiation we can see. Understanding dust dynamics as separate from the gas is a necessary prerequisite to capture the thermodynamic behaviour of a CPD and its environs. In models that presume perfect uniform mixing, the opacity and temperature at any given point in space will be dramatically overestimated or underestimated if the \textit{local} dust-to-gas ratio \textit{at that point} is greater or less than $10^{-2}$. Furthermore, there is no reason to expect dust of different grain sizes to share the same distribution in space, because dust particles of different sizes have different surface-area-to-mass ratios and thus experience different strengths of dust-gas drag.

Previously published work by \citet{BinkertSzulagyi2021} and \citet{SzulagyiBinkert2022} concern three-dimensional hydrodynamical simulations of CPDs with separate gas and dust. The principal differences from this work are as follows: (I) they use only one dust grain size, $a=1\millim$, whereas we allow dust of multiple grain sizes to exist simultaneously, with each dust size possessing its own dynamics; (II) they simulate a larger region of the protoplanetary disc than we do; (III) their simulations are radiative, whereas we adopt a locally isothermal approach; and (IV) they neglect turbulent diffusion of dust, which matters because the main flow feeding the protoplanet is vertical, sourced from far above and below the midplane, and turbulent stirring is what counteracts the gravitational settling of dust which would otherwise pull the dust onto the midplane to form an extremely thin layer. They conclude that planetary gravity vertically stirs the dust, so planet-hosting protoplanetary discs are thicker than expected in the dust and therefore the dust masses of observed protoplanetary discs may be being underestimated.

The purpose of this paper is to assess not only \textit{how much} dust there is in circumplanetary discs but how much dust there is of each grain size. The dust size distribution determines the opacity and, as such, is crucial to understand CPD observations. For example, \citet{Benisty2021} conclude that the circumplanetary disc of PDS 70 c has a dust mass $0.007\Mearth$ if the dust grain size is $1\millim$ or a much more massive $0.031\Mearth$ if the dust grain size is $1\um$. Therefore, we run three-dimensional multifluid hydrodynamical simulations of a circumplanetary disc, covering the gap in which the CPD dwells and the protoplanet within the CPD. Gas and dust are permitted to exist separately, following their separate dynamics, albeit coupled to each other by dust-gas drag. Our approach differs from previous work in that we devote the available computing power to multifluid dust dynamics, rather than to a more sophisticated thermal treatment. We argue that temperature depends so strongly on opacity and thus on the distribution of different-sized dust grains in space that our approach is warranted. In Sect.~\ref{sec:methods}, we lay out the numerical toolset we use, the setup of the simulations and the physical processes they model. Then we give the results of our simulations and compare them to observations in Sect.~\ref{sec:results} and we discuss the implications of these results in Sect.~\ref{sec:discussion}. Finally, our conclusions are offered in Sect.~\ref{sec:conclusions}.

\section{Methods}
\label{sec:methods}

\subsection{Numerical implementation}
\label{sec:numerics}
We run 3D hydrodynamical simulations of a segment of a protoplanetary disc containing a Jupiter-mass protoplanet on a circular orbit at $10\AU$ around a solar-mass star. The orbital radius we use is wider than Jupiter's orbit because we wish to consider protoplanets distant enough to be observable in practice. We use a grid-based Finite-Volume Adaptive Mesh Refinement code called MG \citep{Falle1991, VanLoo2006}. The governing equations for the gas are:
\begin{equation}
\partdiff{\rho_g}{t} + \grad \ve{.} \left( \rho_g \ve{v}_g \right) = 0
\label{eq:gas_mass} \end{equation}
\begin{equation}
\begin{split}
\partdiff{\left(\rho_g \ve{v}_g\right)}{t} +
\grad \ve{.} \left( \rho_g \ve{v}_g \otimes \ve{v}_g - \matri{\sigma} \right)
= -\sum_{i=1}^n \ve{F}_{D,i} - \rho_g \grad \Phi \\
-\rho_g \ve{\Omega}_c \times \left(\ve{\Omega}_c \times \ve{r}\right) - 2\rho_g\ve{\Omega}_c \times \ve{v}_g
\end{split}
\label{eq:gas_momentum} \end{equation}
where $\rho_g$ is the gas density, $\ve{v}_g$ the gas velocity, $\Phi$ the gravitational potential, $n$ the number of different dust species, $\ve{F}_{D,i}$ the drag force by the gas on the $i$\ordinal{th} dust species, and $\ve{\Omega}_c = \Omega_c \unitvector{z}$ the corotation vector with $\Omega_c$ the corotation frequency. In our simulations, we choose to use a frame corotating at frequency $\Omega_c = \sqrt{G\Mstar/ \apl^{3}}$ to keep the protoplanet stationary where $\Mstar$ is the star's mass and $\apl$ the orbital radius of the protoplanet around the star. The position $\ve{r}$ is defined relative to the star which is at the origin, i.e. $\ve{r} = \ve{0}$. The turbulent-viscous stress tensor, $\matri{\sigma}$, is defined as follows:
\begin{equation}
\matri{\sigma} = \scr{\eta}{turb} \left( \grad \otimes \ve{v}_g + \left( \grad \otimes \ve{v}_g \right)^{\mathrm{T}} \right)
- \left(\frac{2}{3} \scr{\eta}{turb} \grad \ve{. v}_g + P\right) \matri{I}
\label{eq:stress_tensor} \end{equation}
where $P$ is the pressure, $\scr{\eta}{turb}$ the turbulent viscosity and $\matri{I}$ the identity matrix. The code is locally isothermal, for computational efficiency. See Sect.~\ref{sec:temperature} for the temperature description. For the $i$\ordinal{th} dust species, the governing equations are:
\begin{equation}
\partdiff{\rho_i}{t} + \grad \ve{.} \left( \rho_i \ve{v}_i - \scr{\eta}{turb} \grad \left( \frac{\rho_i}{\rho_g} \right) \right) = 0
\label{eq:dust_mass} \end{equation}
\begin{equation}
\begin{split}
\partdiff{\left(\rho_i \ve{v}_i\right)}{t} + \grad \ve{.} \left( \rho_i \ve{v}_i \otimes \ve{v}_i
- \left( \scr{\eta}{turb} \grad \left( \frac{\rho_i}{\rho_g} \right) \right)
\otimes \ve{v}_i \right) =
\\
\ve{F}_{D,i} - \rho_i \grad \Phi -
\rho_i \ve{\Omega}_c \times \left(\ve{\Omega}_c \times \ve{r}\right) - 2\rho_i \ve{\Omega}_c \times \ve{v}_i
\end{split}
\label{eq:dust_momentum} \end{equation}
where $\rho_i$ and $\ve{v}_i$ are the density and velocity of the $i$\ordinal{th} dust species, and all other variables are defined above \citep{Morfill1984}. Each dust species is treated as a pressureless fluid.

The MG code uses a Godunov method which is 2\ordinal{nd} order in space and time. For the gas we use a Kurganov-Tadmor Riemann solver, while for the dust Riemann solver we implement the algorithm of \citet{Paardekooper2006}. The dust species are coupled to the gas by dust-gas drag; see Sect.~\ref{sec:drag}. Once the drag coefficients have been calculated, our code uses the algorithm of \citet{BenitezLlambay2019} to solve the effects of dust-gas drag upon all of the dust species and the gas, at once. It solves equations of the form $\ve{F}_{D,i} = - \sum_j \beta_{ij} \left( \ve{v}_i - \ve{v}_j \right)$ by a backward-in-time, implicit, linear-algebra method. Our version of MG is able to simulate an arbitrary number of dust species coexisting with gas, rather than just gas; to work in a corotating frame; and to have protoplanets which exert gravity, accrete matter, and provide heat to their surroundings.

We run one gas-only simulation, four single grain size simulations with gas and one dust species at a time, and one multiple grain size simulation with gas and four dust species simultaneously, with quarter-annulus geometry, and we run one multiple grain size simulation (gas + 4 dust) with full-annulus geometry. The dust grain sizes are $a=1\um$, $10\um$, $100\um$ and $1\millim$. These grain sizes are consistent with the \citet{Miley2021} protoplanetary disc models that produced our initial conditions, as described in Sect.~\ref{sec:initial_conditions}. We obtain the other grain sizes from our choice to use logarithmically even spacing with a factor of 10, in order to explore the behaviour of dust of a wide range of orders of magnitude.

\subsection{Temperature}
\label{sec:temperature}
The disc is assumed to be locally isothermal, with an ideal gas equation of state:
\begin{equation}
P = \frac{\rho_g}{ \Bar{\mu} m_p } k_B T
\label{eq:ideal_gas} \end{equation}
where $\Bar{\mu}$ is mean molecular mass, $m_p$ the mass of a proton, $k_B$ Boltzmann's constant and $T$ temperature.

The initial conditions give the temperature of the unperturbed protoplanetary disc at every point in space, as Sect.~\ref{sec:initial_conditions} explains. The temperature at every point in space is kept equal to its value in these initial conditions, unless the point is close enough that the luminosity of the hot young protoplanet dominates.
\begin{equation}
\func{T}{\ve{r}} = \func{\mathrm{max}}{
\left( \frac{\scr{L}{pl}}{4\pi\left( \func{\scr{d}{pl}}{\ve{r}} \right)^2 \times \sigma} \right)^{1/4}
, \; \; \func{\scr{T}{init}}{\ve{r}} }
\label{eq:heating} \end{equation}
where $\scr{L}{pl}$ is the protoplanet's luminosity, and $\func{\scr{d}{pl}}{\ve{r}}$ the distance from protoplanet is given by $\func{\scr{d}{pl}}{\ve{r}} = \func{\mathrm{max}}{\left|\ve{r} - \scr{\ve{r}}{pl}\right|, \scr{R}{eff}}$. $\scr{R}{eff}$ the ``effective radius'' serves to avoid a singularity at the location of the protoplanet. In this paper we set $\scr{R}{eff}$ to be 8 times the radius of Jupiter. For the protoplanet's luminosity, we use $\scr{L}{pl} = 5.96 \times 10^{-5} \Lsun$, which comes from the equation $\scr{L}{pl} = 4\pi \Rpl^2 \times \sigma \scr{T}{surf}^4$ for a Jupiter-radius protoplanet with the same surface temperature $1600\kelvin$ that was observed by \citet{Christiaens2019} for the giant protoplanet PDS 70 b. This is not expected to be exact for all giant protoplanets but should be of the right order of magnitude.

\subsection{Gravity and accretion}
\label{sec:grav_accretion}
While the self-gravity of the disc material upon itself is neglected, the gravitational acceleration -- which is the same for the gas and dust -- can be straightforwardly calculated using
\begin{equation}
\grad \Phi (\ve{r}) = \frac{G\Mpl  \left( \ve{r} - \scr{\ve{r}}{pl} \right)}
{ \left( \left|\ve{r} - \scr{\ve{r}}{pl}\right|^2 + \left(2 \scr{R}{eff}\right)^2 \right)^{3/2} }
+ \frac{G\Mstar}{\left|\ve{r}\right|^3} \ve{r} + \frac{G\Mpl}{\left| \scr{\ve{r}}{pl} \right|^3} \scr{\ve{r}}{pl}
\label{eq:smoothing} \end{equation}
Note that the first two terms arise directly from the gravitational potentials of the star and protoplanet, while the last term is an indirect, fictitious acceleration due to the gravitational pull of the protoplanet on the star. This is included because our choice of reference frame is keeping the star always at $\ve{r} = \ve{0}$. Furthermore, the direct term of the protoplanet's gravity is artificially smoothed close to the protoplanet using a smoothing radius of $2 \scr{R}{eff}$.

For Eq.~\ref{eq:smoothing}, the protoplanet's mass is fixed at $1\Mjup$ throughout the simulations. These simulations are intended to capture the CPD instantaneously, not to simulate its entire lifetime, which would be computationally prohibitive for high-resolution 3D hydrodynamical simulations.

Accretion has a major impact on the results because the mass budget of the circumplanetary disc is governed by input and output: the flow of mass from the parent protoplanetary disc, and the accretion of mass from the circumplanetary disc onto the protoplanet. Therefore, even though accretion happens on length-scales much smaller than every other length-scale in the problem, it still must be treated with great care. The accretion algorithm we use is Gaussian with distance from the protoplanet and near-linear with time. The mass accreted from a cell in a timestep of length $\Delta t$ is proportional to $\left(1 - \func{\mathrm{exp}}{-\Delta t / \scr{t}{acc}}\right)$ where $\scr{t}{acc}$ is an accretion timescale based on freefall. However, as close to the protoplanet $\Delta t / \scr{t}{acc} \ll 1$, it is in effect linear. For the algorithmic details of the accretion treatment in this paper, see Appendix~\ref{sec:accretion}.

\subsection{Turbulence}
\label{sec:turbulence}
Turbulence in protoplanetary discs is a source of angular momentum transport and can be treated like a viscosity where the kinematic viscosity is given by $\scr{\nu}{turb} = \alpha c_{s,iso}^2 \OmegaK^{-1}$ \citep{ShakuraSunyaev1973} where $c_{s,iso} = \sqrt{P / \rho_g}$ is the isothermal sound speed and $\OmegaK = \sqrt{G\Mstar/R^{3}}$ is the Keplerian frequency. This equation is used to calculate the dynamic viscosity $\scr{\eta}{turb} = \rho_g \scr{\nu}{turb}$ throughout this paper, for both gas and dust. $\scr{\nu}{turb}$ in our simulations is not time-variable; it is calculated using the initial conditions. The turbulent/diffusive terms in Eqs~\ref{eq:dust_mass} and \ref{eq:dust_momentum} are for turbulent stirring, which must not be neglected because the balance between it and the settling due to drag and gravity sets the scale height for dust of each grain size \citep{Youdin2007}. Without it, the dust would settle into a super-dense, gravitationally unstable layer at the midplane \citep{Goldreich1973}.

Observationally, the general consensus is that $\alpha$ is around $10^{-4}$ - $10^{-3}$ in Class II discs: for instance, \citet{Pinte2016} look at the continuum emission of the disc HL Tau and, by modelling the vertical settling of dust, they deduce an $\alpha$ of order a few times $10^{-4}$. With a different method, \citet{Trapman2020} analyse protoplanetary discs' viscous spreading by comparing PPDs' ages to their outer radii for a sample in the Lupus star-forming region and they conclude that $\alpha$ is generally in the $10^{-4}$ - $10^{-3}$ range. In this paper we take $\alpha = 10^{-3}$. We choose the upper end of the $10^{-4}$ - $10^{-3}$ range because higher $\alpha$ means higher dust scale heights, which are less computationally expensive to capture.

\subsection{Dust-gas drag}
\label{sec:drag}
Dust-gas drag is treated as one of two regimes, depending on comparing the dust grain size $a$ to the mean free path $\lambda$ of the gas: the Epstein drag regime when $a \leq \frac{9}{4} \lambda$ and the Stokes regime when $a > \frac{9}{4} \lambda$. The mean free path can be expressed in terms of the gas density and collisional cross section $\scr{\sigma}{coll}$ as $\lambda = \Bar{\mu} m_p / \left(\rho_g \scr{\sigma}{coll}\right)$ with $\scr{\sigma}{coll}$ taken to be $2 \times 10^{-19} \metre^2$ and the mean molecular mass $\Bar{\mu}=2.3$ \citep{Dipierro2018}. Following \citet{Dipierro2018}, we use the following drag equations:
\begin{equation}
\ve{F}_{D,i} = -\frac{\rho_g v_{th}}{\rho_m a}\rho_i \left(\ve{v}_i - \ve{v}_g \right) \times
\begin{cases}
1 & \text{if $a \leq \frac{9}{4} \lambda$ (Epstein)}\\
\frac{9 \lambda}{4 a} & \text{if $a > \frac{9}{4} \lambda$ (Stokes)}
\end{cases}
\label{eq:dustgasdrag} \end{equation}
where $\rho_m$ is the material density of a dust grain which we take to be $3000\kg \metre^{-3}$, $v_{th}$ the thermal speed which is $v_{th} = c_{s,iso} \sqrt{8 / \pi}$ in the Boltzmann distribution. Of course, the $i$\ordinal{th} dust species exerts an equal and opposite drag force on the gas: $\ve{F}_{D,g} = - \Sigma_{i=1}^n \ve{F}_{D,i}$.

\subsection{Initial and boundary conditions}
\label{sec:initial_conditions}
The simulations are done in 3D cylindrical polar coordinates $\left(R, \phi, z\right)$ in a stellar-centric frame. That is, the star is always at $\ve{r} = \ve{0}$. Computational units are chosen so that $\apl$, the radius of the protoplanet's orbit around the star, is 1 and the period of the protoplanet's orbit is also 1. Thus, in this corotating frame, the protoplanet is always at $R=1$, $\phi=0$, $z=0$. The region we simulate is $0.7 \leq R \leq 1.3$, $0 \leq z \leq 0.2$. For every simulation but the last, the simulated region is $\frac{-1}{4}\pi \leq \phi \leq \frac{1}{4}\pi$, one quarter of an annulus. The $\phi$ (azimuthal) boundary conditions are periodic, so that information is not lost as matter orbits the star, following \citet{AyliffeBate2009A, AyliffeBate2009B}. For the final simulation, we simulate the full annulus, $2\pi\radians$, at the same resolution as was done for the quarter-annulus. The simulations only include the upper half of the disc because mirror-symmetry at the midplane is assumed. The upper $z$ (vertical) boundary condition and both of the $R$ (radial) boundary conditions are fixed at their values from the initial conditions described below.

These are multi-resolution simulations, with the higher-resolution levels existing only in the vicinity of the protoplanet. The low-resolution level which covers the entire grid, Level 1, has resolution 120 in $R$, 40 in $z$ and 316 in $\phi$ (for quarter-annulus). That yields cells of size $\sim 0.005\apl$ in all three dimensions. The accretion near to the protoplanet takes place on length-scales $\sim$ the radius of Jupiter, $5 \times 10^{-4}\AU$. Using such high resolution for the entire simulation is prohibited by computation time. Therefore we use a static mesh refinement. If a Level-1 cell is within 512 Jupiter radii of the protoplanet, it is divided into 8 Level-2 cells. If one of these Level-2 cells then lies within 256 Jupiter radii, it is further divided into 8 Level-3 cells, and so on. The base grid is fully resolved and the highest grid level is 6, so that our maximum resolution is $2^5$ times the base grid's resolution.

Initial conditions and boundary conditions for the simulations come from star+protoplanetary disc models developed by \citet{Miley2021}. These models use the Monte Carlo radiative transfer code \textsc{mcmax} \citep{Min2009} to produce self-consistent 2D solutions for temperature and densities in an axisymmetric protoplanetary disc. The parameters of the \citet{Miley2021} model that we use are shown in Table~\ref{tab:params}.

\begin{table}
\centering
\begin{tabular}{lcr} % alignment of columns
\hline
Parameter & Value & Units\\
\hline
Stellar mass & $1$ & $\Msun$\\
Mass of protoplanetary disc & $0.05$ & $\Msun$\\
Age of protoplanetary disc & $1 \times 10^6$ & $\yr$\\
Dust size distribution power-law & $-3.5$ & \\
Minimum dust grain size & $1 \times 10^{-8}$ & $\metre$\\
Maximum dust grain size & $1 \times 10^{-3}$ & $\metre$\\
Turbulent alpha parameter & $1 \times 10^{-3}$ & \\
\hline
\end{tabular}
\caption{Input parameters for the \citet{Miley2021} models that we used to generate initial and boundary conditions for our simulations.}
\label{tab:params} \end{table}

The \citet{Miley2021} models are static. We have taken from the models the temperature, gas density, and total dust density summed over all grain sizes: $\func{T}{R,z}$, $\func{\rho_g}{R,z}$, $\func{\scr{\rho}{all\;dust}}{R,z}$. For velocities, we initially approximate as follows: $v_R = v_z = 0$ for both gas and dust; $v_\phi = \sqrt{G\Mstar R^{-1} - 3 P \rho_g^{-1}}$ for gas; and $v_\phi = \sqrt{G\Mstar R^{-1}}$ for dust. This is a simplified form of an analytical protoplanetary disc expression; see Eq. 13 of \citet{Nelson2013} and remove the $q\left(1 - R / \sqrt{R^2 + z^2}\right)$ term for simplicity. We set $p$ and $q$, the power-law indices for the dependence of midplane gas density and temperature (respectively) on $R$, to $p=-2.5$ and $q-0.5$. Dust, being pressureless, lacks the gas's $\left(p + q\right)\left(H/R\right)^2$ term. Since the initial star+disc models span the whole protoplanetary disc from $0.24\AU \leq R \leq 200\AU$, their grid is much coarser in space than this paper. That necessitates logarithmic interpolation to convert them to appropriate initial and boundary conditions. Thus, taking temperatures and densities from the initial star+disc models and velocities from an approximate analytical prescription, we obtain $\{T,\rho,v_R,v_\phi,v_z\}$ as a function of $R$ and $z$.

While this provides us the initial conditions for the gas-only and single-grain models with gas (where we assume that the grain size is either i.e. $1\um$, $10\um$, $100\um$ or $1\millim$), it does not directly give us the dust distribution for the multiple grain size simulations. In the multiple grain size simulations, the dust mass is divided between four grain sizes and we must obtain the density for each individual grain size $\rho_i$ from the overall summed dust density. We assume that these grain sizes, i.e. $\bar{a}_1 = 1\um$, $\bar{a}_2 = 10\um$, $\bar{a}_3 = 100\um$ and $\bar{a}_4 =1\millim$, are representative of a continuous grain size distribution given by $\diff{\func{N}{a}}{a} = N_0 a^{-3.5}$ where $N_0$ is a normalisation factor \citep{Mathis1977}. In principle the mass density for each grain radius can be calculated using 
\begin{equation}
\rho_i = \func{m}{a_i} \left. \diff{\func{N}{a}}{a} \right|_{a_i} = \frac{4\pi\rho_m}{3} N_0 a_i^{-0.5}
\end{equation}
The normalisation factor $N_0$ would then be determined by summing the mass densities and setting this sum equal to the total dust density. However, such an approach ignores the fact that the given grain sizes represent a range of grain radii with $\bar{a}_i \in [a_i, a_{i+1}]$. A meaningful choice for a characteristic grain size is such that both the number and mass density of the bin can be reproduced simultaneously. This requires 
\begin{equation}
\frac{4\pi\rho_m}{3} \bar{a}_i^3 = \frac{ \func{M}{a_i, a_{i+1}} }{ \func{N}{a_i, a_{i+1}} }
\end{equation}
where $N(a_i, a_{i+1})$ and $M(a_i, a_{i+1})$ are the total number density and mass density, respectively, of grains with radii between $a_i$ and $a_{i+1}$.
\begin{equation}
\func{M}{a_i, a_{i+1}} = \int_{a_i}^{a_{i+1}} \func{m}{a} \diff{\func{N}{a}}{a} \ \mathrm{d}a
\end{equation}
With our choice of characteristic grain radii $\bar{a}_i$, this actually sets the lower and upper limit of each grain size bin, i.e. $a_i \approx 0.4517 \bar{a}_i$, while $a_{i+1} \approx 4.517 \bar{a}_i$. Using these limits we can then calculate the mass densities and determine the normalisation factor $N_0$, and thus $\rho_i = M(a_i, a_{i+1})$. This method is applied at every point in space.

However, as Sect.~\ref{sec:introduction} elaborates, the dust grain size distribution is observably not the same everywhere in space. Hence, the initial and boundary conditions from the above procedure are only provisional. To obtain our true initial and boundary conditions, we take the provisional $\{T,\rho,v_R,v_\phi,v_z\} \left(R,z\right)$ values and we plug them into the MG hydrodynamics code, now simulating a slightly larger region: $0.65 \leq R \leq 1.35$, $0 \leq z \leq 0.22$. This protoplanetary disc is then allowed to evolve freely for 10 orbital periods, with all the same physics except that axisymmetry is assumed and no protoplanets are present. This serves to ``relax'' the values from the initial star+disc models to a stable steady state, prior to the implantation of protoplanets. During this relaxation phase, the dust settles to the scale height appropriate for its grain size, except at the boundaries where the boundary conditions are pinned to the initial conditions. For this reason we use a larger simulated region during relaxation which prevents any distortion near the boundaries from entering the main simulations. Furthermore it produces a flux of inward radial-drifting dust. It will not perfectly capture the phenomenon of radial drift because that takes place on timescales of order the disc lifetime, which greatly exceeds the length of these simulations. The resultant relaxed, steady-state, fully hydrodynamic models are used as the initial and boundary conditions for the main simulations.

\subsection{Implanting protoplanets}
\label{sec:implantation}
Protoplanet growth during the runaway gas accretion phase takes place on timescales $\sim 10^4 - 10^6 \yr$ \citep{Helled2014}. For contrast, the relevant dynamical timescale of our simulations is the orbital period, which is $\sim 30 \yr$ at $\apl \sim 10 \AU$ around a star of mass $\sim 1 \Msun$. The timescale of protoplanet growth is so many orders of magnitude longer than the timescale of our simulations that protoplanet growth is effectively static on our timescales. Thus, for our simulations to be accurate, we need them to have settled into a quasi-static state.

Numerical breakdown would be caused by instantaneous insertion of a Jupiter-mass protoplanet into an unperturbed protoplanetary disc model. To avoid this, the protoplanet's mass is set to $\Mpl=0$ at $t=0$ and it is linearly grown to its desired mass over the first 3 orbital periods of the main simulation. For our simulations the desired mass is $1\Mjup$. This super-fast linear growth is not a representation of the planet formation process but purely a tool to avert numerical breakdown.

The super-fast protoplanet implantation excites the protoplanetary disc to a temporary unsustainable state with extremely large amounts of matter clustering around the protoplanet and thus extremely high accretion rates. Therefore, even though the protoplanet is at full mass at $t=3$ orbits, a snapshot of the simulation at $t=3$ orbits is not conclusive. It is necessary to give the simulation more time to allow it to relax into a sustainable steady state. How much time, and how we determine that, is discussed in Sect.~\ref{sec:results}.

\section{Results}
\label{sec:results}
\subsection{Gas dynamics}
\label{sec:gas}

\begin{figure}
\includegraphics[width=\columnwidth]{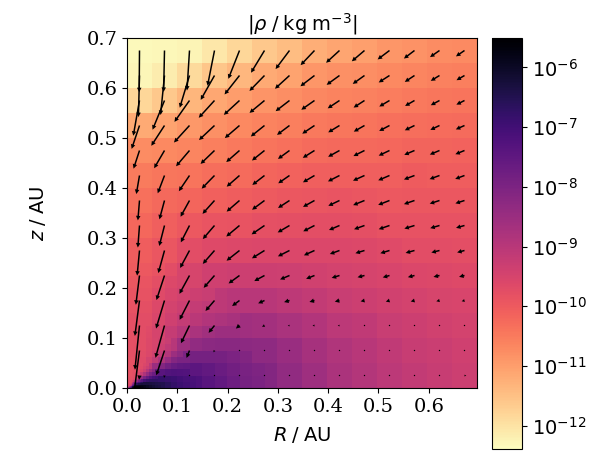}
\caption{The density distribution of a circumplanetary disc, in a frame comoving with the protoplanet, in a gas-only simulation. The protoplanet is at (0,0). $R$ and $z$ are measured from the protoplanet. The densities and velocities presented here have been mass-averaged across $\phi$, the azimuthal coordinate from the protoplanet. The arrows show the mass-averaged velocity vectors, or rather their $R$ and $z$ components. The $\phi$ component of velocity, orbiting around the protoplanet, is not shown.}
\label{fig:circumplanetary} \end{figure}

\begin{figure}
\includegraphics[width=\columnwidth]{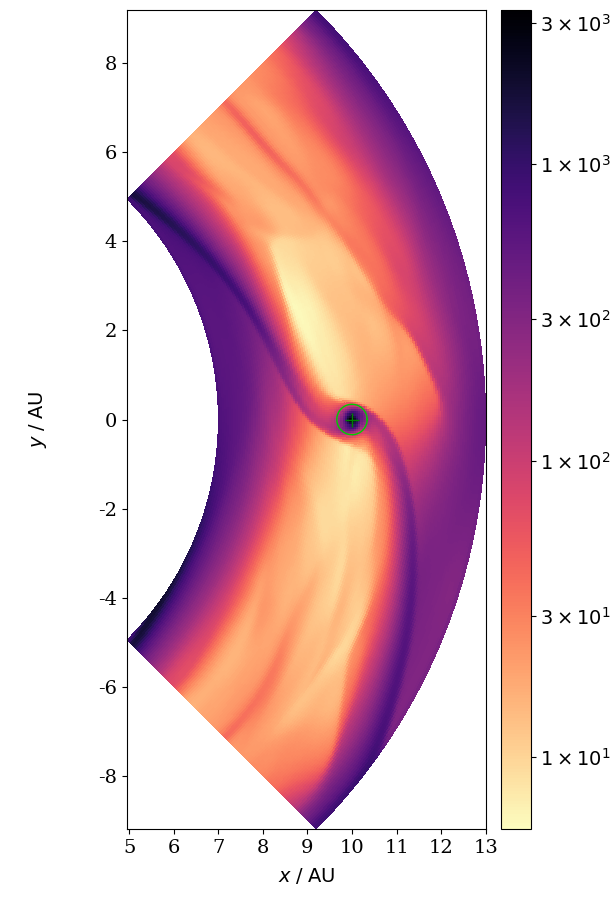}
\caption{Gas surface density in units $\kg \metre^{-2}$ for the gas-only simulation at $t = 50$ orbits after the implantation of the protoplanet. The green semicircle denotes a distance of $0.5\RHill$ from the protoplanet, which is marked with a cross.}
\label{fig:sigma_gasonly} \end{figure}

First we start with a gas-only simulation. It serves as a fiduciary model to confirm that our code is working as it should, reproducing the opening of a gap and the formation of a circumplanetary disc seen in previous studies \citep[e.g.][]{Kley1999, Nelson2000, Machida2008}.  Fig.~\ref{fig:sigma_gasonly} shows the gas surface density 50 orbits after the protoplanet was introduced in the numerical domain. The tidal torques exerted by the protoplanet indeed perturb the disc gas density in the form of trailing spiral shock waves. These open up an annular gap in the disc, although, after 50 orbits, the disc is not yet fully cleared and some disc material on a co-rotating orbit with the protoplanet is still present. This gas oscillates on horseshoe-shaped orbits in the frame corotating with the protoplanet. Henceforth we refer to this as the horseshoe region. 

Simultaneously a CPD forms around the protoplanet. Fig.~\ref{fig:circumplanetary} shows the azimuthally averaged density distribution within one Hill radius, $\RHill = 0.69\AU$ and shows a flared disc structure which is notably denser than its surrounding material. The disc itself is rotationally supported, while additional gas is fed to the CPD by meridional flows \citep[as seen in e.g.][]{SzulagyiMasset2016}. The CPD extends to a distance of about $\approx 0.5 \RHill$ from the protoplanet corresponding roughly to the extent of protoplanet's Roche lobe. Therefore, throughout this paper, we define the CPD mass as twice the mass in all cells within a distance $\leq 0.5 \RHill$ of the protoplanet. Note that the factor of 2 is because we use symmetry boundary conditions at the midplane.

\begin{figure}
\includegraphics[width=\columnwidth]{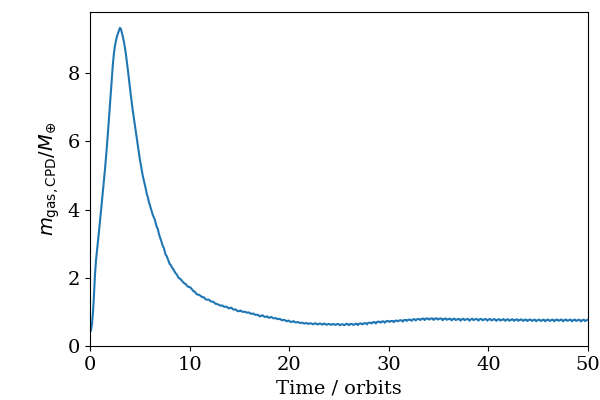}
\caption{The mass of a CPD in a gas-only simulation, over time. The CPD is settling into a steady state after the implantation of a protoplanet to the parent protoplanetary disc at $t=0$. This metric serves to inform us of when the CPD has reached a steady state.}
\label{fig:steadystate} \end{figure}

As previously stated (Sect.~\ref{sec:implantation}), after the implantation of the protoplanet, the system requires some time to settle down. Fig.~\ref{fig:steadystate} shows the temporal evolution of the gas mass in the CPD in the gas-only simulation. It shows a rapid increase in the CPD gas mass which reaches a maximum after 3 orbits. Then the CPD mass reduces as more gas is accreted by the protoplanet than is deposited on the CPD. After about 20-25 orbits the CPD mass loss and gain balance each other and the CPD gas mass remains constant at about $0.76\Mearth$. The extreme clustering of matter near the protoplanet in the early part of these simulations is a numerical artefact due to  the super-fast implantation of the protoplanet between $t=0$ and $t=3$ orbits. Only some time after the implantation phase (about 20-30 orbits), the CPD reaches a quasi-steady state, and it is this state that we analyse in this paper. This does not mean the system does not continue to evolve. Two-dimensional simulations of the late-time behaviour shows that the gap first becomes devoid of gas and that subsequently the inner disc (between star and protoplanet) disappears as the gas is accreted by the star \citep{Nelson2000}. However, computational restrictions of high-resolution three-dimensional simulations do not allow us to follow the CPD evolution up to such long timescales.

\begin{figure*}
\includegraphics[width=\textwidth]{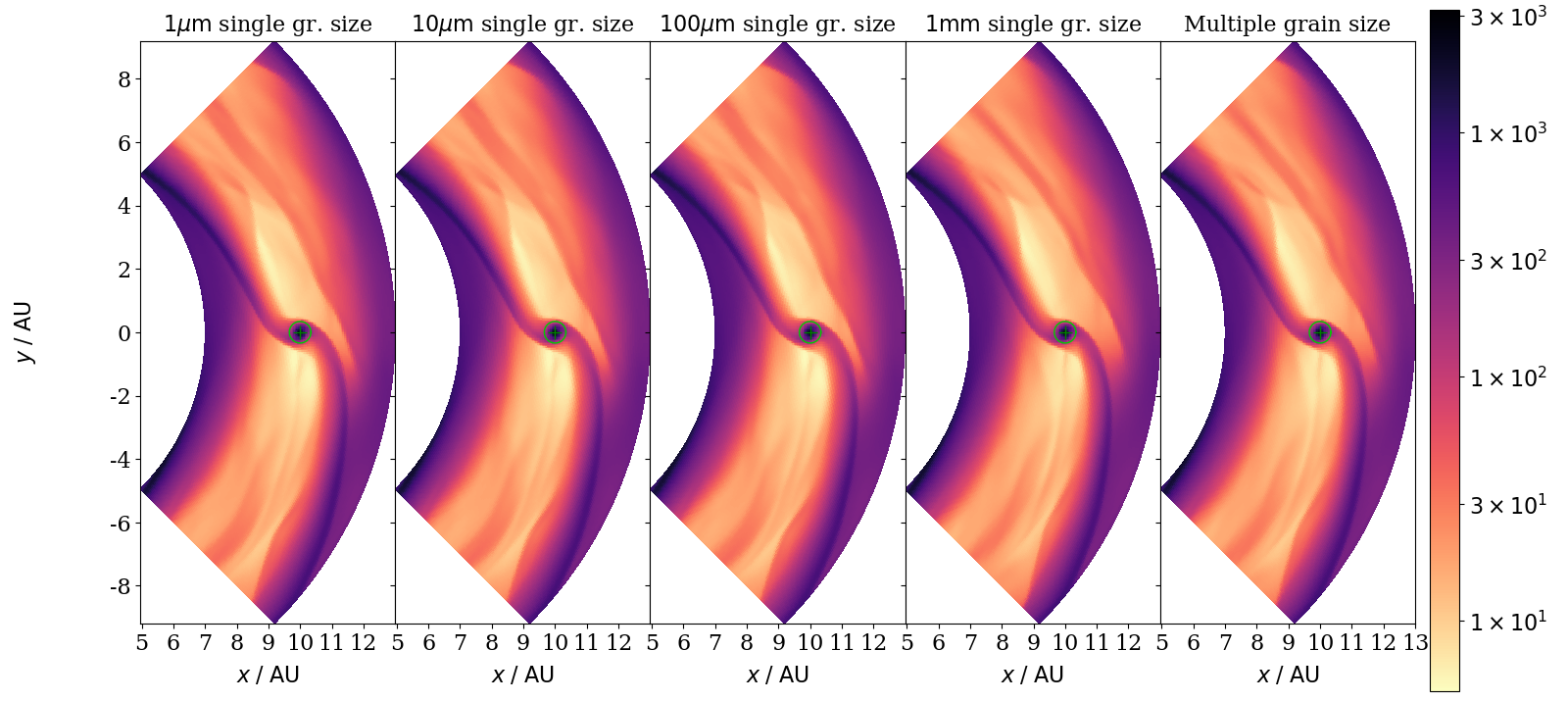}
\caption{Surface density of gas in protoplanetary discs, in units $\kg \metre^{-2}$, after $t=50$ orbits since the implantation of the protoplanet. From left to right, the subplots show the single grain size simulations for $a=1\um$, $10\um$, $100\um$ and $1\millim$ and finally the (quarter-annulus) multiple grain size simulation on the far right. The green semicircle denotes a distance of $0.5\RHill$ from the protoplanet, which is marked with a cross.}
\label{fig:sigmagas_singleandmulti} \end{figure*}

\begin{figure}
\includegraphics[width=8.4cm]{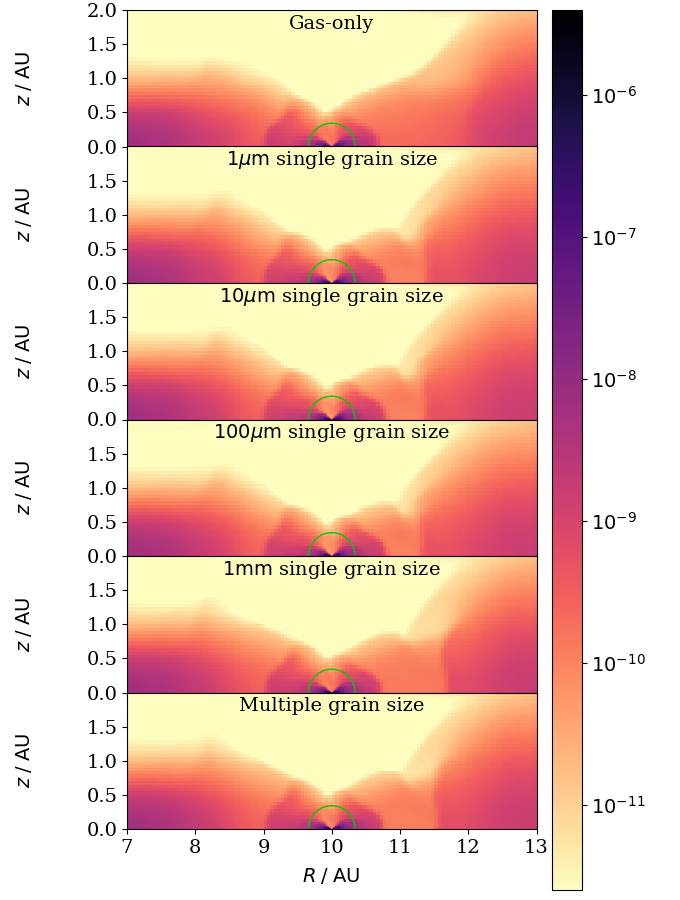}
\caption{Vertical slice at $\phi = 0$ of the gas density, in units $\kg \metre^{-3}$, after $t=50$ orbits since the implantation of the protoplanet. From top to bottom, the subplots show the gas-only simulation, then the single grain size simulations with $a=1\um$, $10\um$, $100\um$, and $1\millim$, and then the multiple grain size simulation (all quarter-annulus). The green semicircle denotes the distance of $0.5 \RHill$ from the protoplanet.}
\label{fig:rhogas_verticalcut} \end{figure}

\begin{figure}
\includegraphics[width=8.4cm]{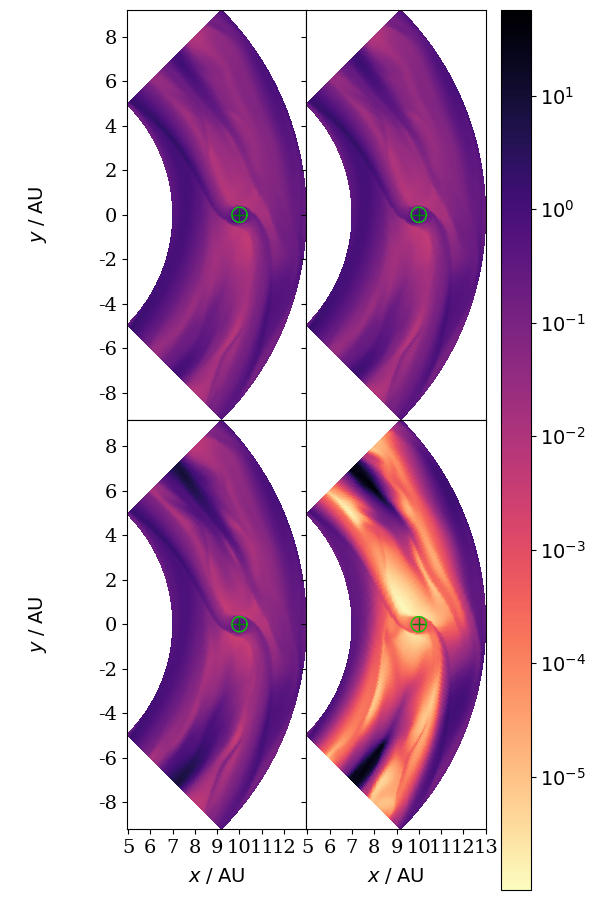}
\caption{Dust surface density for the single grain size simulations, i.e. $a=1\um$ (top left), $10\um$ (top right), $100\um$ (bottom left), and $1\millim$ (bottom right) in $\kg \metre^{-2}$ after $t=50$ orbits since the implantation of the protoplanet. The green semicircle denotes a distance of $0.5\RHill$ from the protoplanet, which is marked with a cross.}
\label{fig:sigmadust} \end{figure}

\subsection{Single grain size dynamics}
\label{sec:singlegrainsize}
Now we include single-sized dust grains as well as the gas.
Fig.~\ref{fig:sigmagas_singleandmulti} shows the gas's surface density (density integrated along the $z$ axis) and Fig.~\ref{fig:rhogas_verticalcut} the density of a slice at $\phi=0$. From these it is clear that the general effect of the dust grains on the gas structure is small, i.e. the width of the annular gap in the disc and the structure of the spiral arms connecting the CPD with the disc do not change at all. The main differences are seen in the structure of the horseshoe region: its location and thickness differs compared to the gas-only simulation and even between the single grain size simulations.

To interpret this we need to understand the interaction between the gas and dust grains. In a general situation when the dust density is much smaller than the gas density, the radial motion of the dust particles is given by \citep{Dipierro2018, Zhu2012}
\begin{equation}
v_{d,R} = \frac{v_{g,R} \mathrm{St}^{-1} + v_p}{\mathrm{St} + \mathrm{St}^{-1}} + \frac{ \scr{v}{visc} }{1+\mathrm{St}^2} - \frac{\scr{\eta}{turb}}{\rho_d} \partdiff{\;}{R}
\left(\frac{\rho_d}{\rho_g}\right)
\label{eq:theory_dustveloc} \end{equation}
where $\mathrm{St} = \frac{\rho_m a}{\rho_g v_{th}}\OmegaK$ is the Stokes number of the dust grains, $v_p = \frac{1}{\rho_g\OmegaK} \partdiff{P}{R}$ the typical dust drift velocity due to pressure differences and $\scr{v}{visc} = \frac{2}{\rho_g \OmegaK} \divF{\matri{\sigma}}|_\phi$ the radial drift due to viscous torques. The last term is the drift due to dust diffusion. For low Stokes numbers the dust grains closely follow the gas: the gas-grain drag dominates and the viscous drift is small compared to the gas velocity. However, when there is a large gradient in the dust-to-gas mass ratio, dust diffusion can become important. Grains with a high Stokes number, i.e. $\mathrm{St} > 0.1$, decouple from the gas and the drift due to pressure gradients plays a significant role. In our simulations only the $1\millim$ grain model has high enough Stokes numbers for the dust and gas to decouple from each other, although the decoupling transition already starts at the smaller grain size of $100\um$. Fig.~\ref{fig:sigmadust} shows the dust surface density for each single grain size simulation. The surface density structure is nearly identical for $1\um$, $10\um$ and $100\um$, but is significantly different for $1\millim$. Especially the dust density within the annular gap is a few orders of magnitude lower. This is an effect of the pressure-gradient drift, i.e. at the outer edge of the gap a pressure bump forms an effective barrier for the grains to drift inward. As a consequence the gap becomes devoid of $1\millim$ dust grains. This process is referred to as dust filtering \citep{Rice2006} and observed in many simulations \citep[e.g.][]{Zhu2012}. 

\begin{figure}
\includegraphics[width=8.4cm]{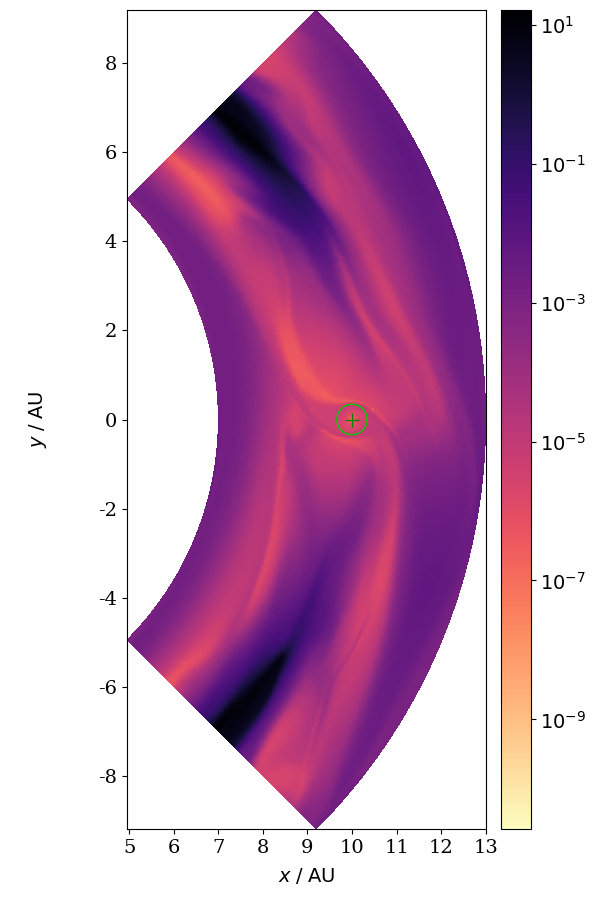}
\caption{Dust-to-gas ratio in the midplane after $t=50$ orbits since the implantation of the protoplanet, for the single grain size simulation of grain size $1\millim$. The green semicircle denotes a distance of $0.5\RHill$ from the protoplanet, which is marked with a cross.}
\label{fig:1mm_dustgasratio} \end{figure}

Another significant difference seen for the $1\millim$ simulation is that, although the gap is devoid of  dust grains, the dust grains in the corotating region are trapped because of pressure gradients. As gas moves out of the annular gap, the dust-to-gas mass ratio therefore increases significantly and dust grains actually become the dominant mass carriers. Fig.~\ref{fig:1mm_dustgasratio} shows that the dust-to-gas mass ratio in the horseshoe region is 3 orders of magnitude larger than for typical ISM values. When this happens the back-reaction (or drag force) of the dust grains on the gas can no longer be neglected. The qualitative analysis of \citet{Dipierro2018} shows that the back-reaction already becomes important when $\rho_d/\rho_g > \alpha/(\mathrm{St} - \alpha)$. However, \citet{Dipierro2018} did not include the effect of dust diffusion. Dust diffusion actually acts earlier as can be seen in the comparison of the thickness of the horseshoe region between the gas-only and single grain size simulations (see Figs.~\ref{fig:sigma_gasonly} and \ref{fig:sigmagas_singleandmulti}). At the boundaries of the horseshoe region the dust-to-gas mass ratio changes rapidly which gives rise to dust diffusion drift and pushes the dust away from the horseshoe region. As the gas and dust are strongly coupled, it actually drags the gas with it. While dust diffusion is also important for the $1\millim$ grains, the gas and dust are only weakly coupled leading to a thin horseshoe region as in the gas-only simulation, but a thick region in the dust. As we mentioned earlier, the dust-gas decoupling can already be noticed in the $100\um$ model, as it shows a dust distribution in between the smallest grain size simulations and the largest grain size simulation.

\begin{figure}
\includegraphics[width=8.4cm]{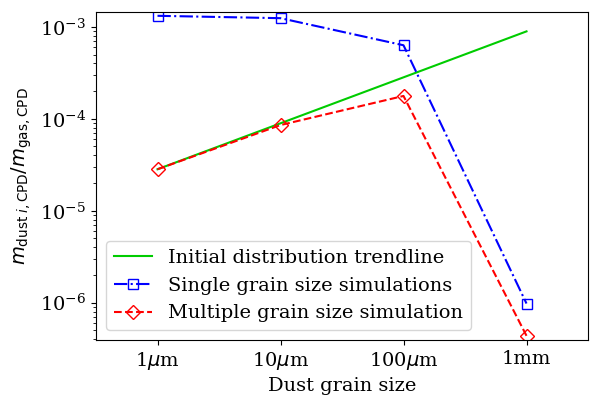}
\caption{Dust-to-gas mass ratio of the circumplanetary disc at $t=50$ orbits for single grain size (blue dash-dotted) and multiple grain size simulation with quarter-annulus (red dashed). The green solid line is a power-law \citet{Mathis1977} distribution normalised with the value at $1\um$.}
\label{fig:dgratioCPDmulti} \end{figure}

\begin{figure}
\includegraphics[width=8.4cm]{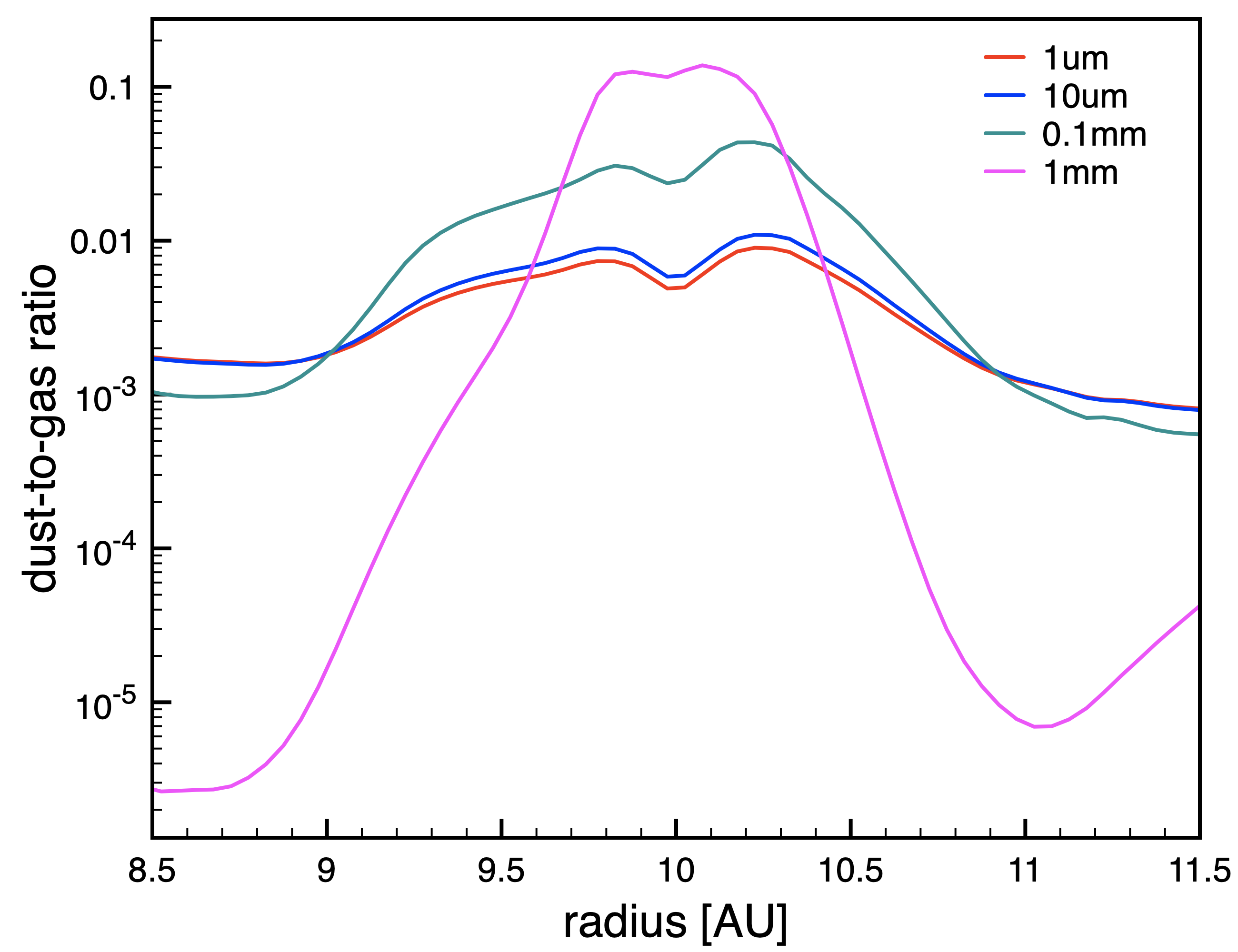}
\caption{Azimuthally and vertically averaged dust-to-gas mass ratio after $t=50$ orbits for the different single grain size simulations.}
\label{fig:dgratiogapsingle} \end{figure}

So, the inclusion of dust grains does not change the gas dynamics, especially not the formation of a CPD around the protoplanet. However, as we have seen, the dynamics of the dust depends on the Stokes number and, thus, the size of the grains. This also has consequences for the dust content of the CPD. As seen in Sect.~\ref{sec:gas}, gas form inside the annular gap is transported to the CPD via meridional flows. As in the $1\millim$ simulation, the gap is devoid of dust grains, it is likely that the CPD has no dust in it either. Fig.~\ref{fig:dgratioCPDmulti} shows the dust-to-gas ratio of the CPD and, indeed, the ratio for the $1\millim$ simulation decreases to $10^{-6}$ while the smaller grain size simulations have equal values around $10^{-3}$ (although for the $100\um$ model it is a factor of 2 lower). Note that the smaller grain size simulations also have a lower dust-to-gas mass ratio than the default value of $10^{-2}$. This is because the pressure maximum at the centre of the gap traps the dust grains to form the horseshoe region. Although not as efficient as in the $1\millim$ simulation, dust in the smaller grain size simulations is still more efficiently trapped than the gas. This is why, as Fig.~\ref{fig:dgratiogapsingle} shows, the dust-to-gas ratio in the horseshoe region is slightly above $10^{-2}$, whereas it is lower elsewhere in the gap. That is also seen in 2D simulations, e.g. \citet{Drazkowska2019}. Actually, the ratio we obtain in the CPD is the same as in the gap, reinforcing the notion that material in the CPD is replenished by meridional flows.

\subsection{Multiple grain size dynamics}
\label{sec:multiplegrainsize}
In the previous section, we studied the behaviour of each dust species separately. The results show that the dust content of the CPD depends directly on the dust content of the annular gap and, thus, is grain size dependent due to dust filtration. Furthermore, the grain size affects the dynamics of the system as grains with a high Stokes number (or large grain size) decouple from the gas and dust-to-gas feedback becomes important. As there is a dust grain size distribution within protoplanetary discs, it is therefore important to consider the dynamics of multiple grain species simultaneously. The large, weakly coupled grains potentially modify the dynamics of the smaller well-coupled grains. 

\begin{figure}
\includegraphics[width=8.4cm]{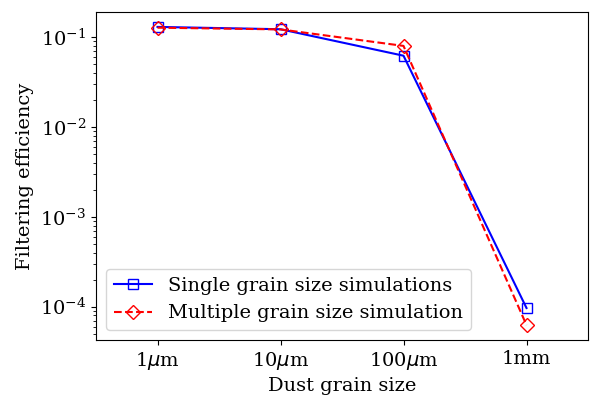}
\caption{Dust-to-gas mass ratio in the CPD normalised to the initial dust-to-gas mass ratio for that grain species. The blue, solid line shows the single size grain simulations while the red, dashed line shows the quarter-annulus multiple grain size simulation.}
\label{fig:filtering_efficiency} \end{figure}

Figs.~\ref{fig:sigmagas_singleandmulti} and \ref{fig:rhogas_verticalcut} show that the gas structure for the quarter-annulus multiple grain size simulation is similar to the $1\millim$ single grain size simulation. This is not surprising as \citet{Dipierro2018} show that, for a continuous dust distribution, the effect of dust-gas drag on both the dust and gas is set by the parameters
\begin{equation}
\lambda_k = \sum_{i=1}^n{\frac{\mathrm{St}_i^k}{1 + \mathrm{St}_i^2} \frac{\rho_i}{\rho_g}}
\label{eq:lambdas} \end{equation}
where $k \in \{0,1\}$ and $n$ is the number of dust grain size bins. For an MRN \citep{Mathis1977} distribution, the value of $\lambda_0$ and $\lambda_1$ are solely determined by the Stokes number of the largest grains. This is because the largest bin (represented by the average bin grain size of $1\millim$) not only has the highest Stokes number, it also contains most of the dust mass. Thus, the dynamics, and thus the structure, of the largest grains and the gas are extremely similar to the single grain size simulation for $1\millim$. The dynamics of the smaller grains that are strongly coupled to the gas does change in relation to their single grain size simulations. The density structure in these smaller grains now looks like the $1\millim$ dust grain structure. As seen in Sect.~\ref{sec:singlegrainsize}, most of the difference is in the horseshoe region and not the CPD. This can be seen from Fig.~\ref{fig:dgratioCPDmulti} which shows the dust-to-gas mass ratio in the CPD. While the $1\um$ and $10\um$ dust grain size bins follow roughly the expected MRN distribution, the dust mass in the $100\um$ bin is a factor of 2 less than would be expected if it followed the MRN distribution, and the $1\millim$ mass is 3 orders of magnitude smaller. Fig.~\ref{fig:filtering_efficiency} shows that the multiple grain size simulation has the same filtering efficiency -- CPD dust-to-gas mass ratio of a dust species, normalised by the initial dust-to-gas ratio of that species -- for the different dust species as in the single grain size simulations. It is thus clear that dust filtering acts in the multiple grain size simulation as it does in the single grain size simulations and that every dust species behaves dynamically as if it and the gas were an isolated system.

An important consequence of the multiple grain size simulation is that, although the CPD is populated with a wide size range of dust grains that are well coupled to the gas, the total dust-to-gas mass ratio of the CPD is much less than in the single size grain simulations, i.e.  $\approx 3\times10^{-4}$ compared to $10^{-3}$. This is because most of the protoplanetary disc dust mass is in the $1\millim$ bin. Dust filtration stops these large dust grains from flowing into the protoplanet-carved gap and, thus, also onto the CPD.

\subsection{Full-annulus geometry}
\label{sec:QCvsC}
\begin{figure}
\includegraphics[width=8.4cm]{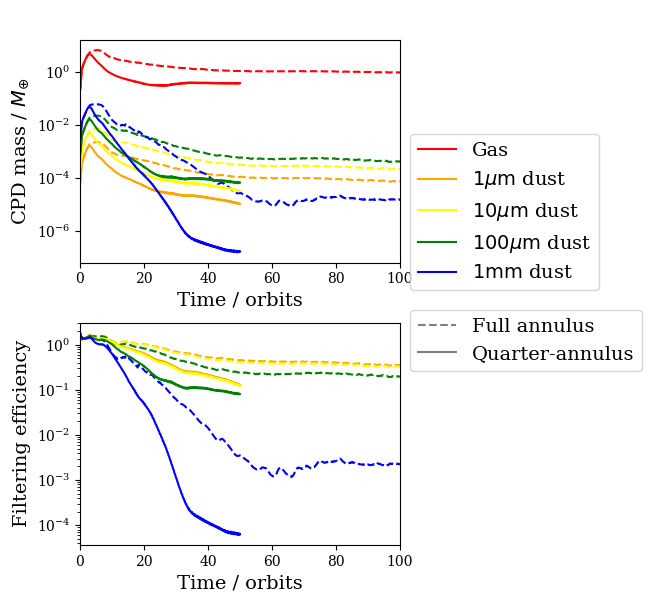}
\caption{Circumplanetary disc masses and filtering efficiencies over time in the quarter-annulus and full-annulus multifluid simulations. Filtering efficiency is as defined in Sect.~\ref{sec:multiplegrainsize}: a dimensionless ratio for each dust species, proportional to that dust species's CPD mass.}
\label{fig:QCvsC_steadystate} \end{figure}

\begin{figure}
\includegraphics[width=8.4cm]{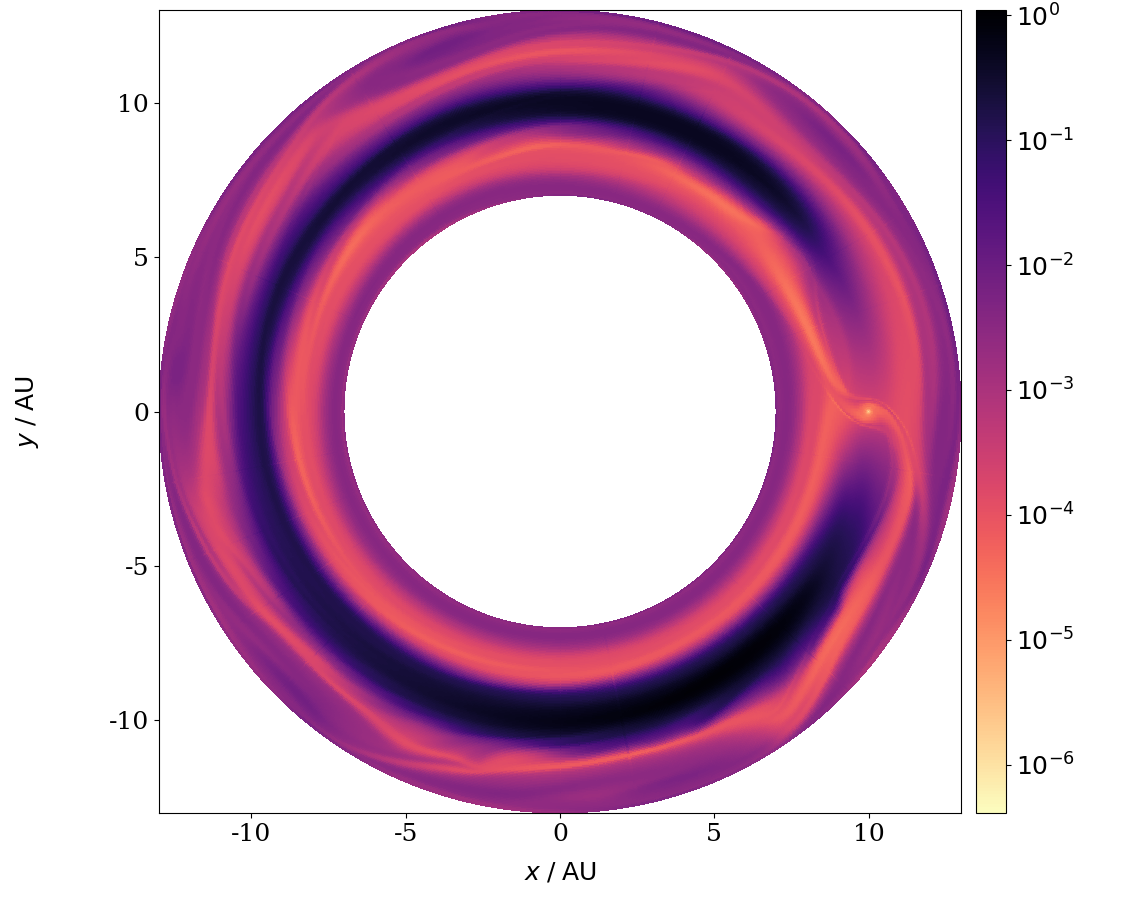}
\caption{Dust-to-gas ratio of the $1\millim$ dust species in the midplane after $t=100$ orbits since the implantation of the protoplanet, for the full-annulus multiple grain size simulation.}
\label{fig:dg_fullcircle} \end{figure}

Our quarter-annulus simulations provide an excellent comparison between the gas-only, single-grain and multiple grain models, but the periodic boundary conditions potentially affect the obtained results. To assess the effects, we run one additional simulation, which is identical in every way to the quarter-annulus multiple grain size simulation from Sect.~\ref{sec:multiplegrainsize} except that it covers the full annulus, i.e. $2\pi\radians$, without loss of resolution. This full-annulus multiple grain size simulation takes longer to settle into steady state than its quarter-annulus counterpart, because it has more mass in the gap. Therefore, we run it for longer up to $t=100$ orbits, not $t=50$ as before. From Fig.~\ref{fig:QCvsC_steadystate} it is apparent that the simulation has reached a quasi-steady state by then. Fig.~\ref{fig:dg_fullcircle} shows that, qualitatively, this full-annulus result does not dramatically differ from the quarter-annulus results as compared to e.g. Fig.~\ref{fig:1mm_dustgasratio}. There are some small local structures in the horseshoe region in Fig.~\ref{fig:1mm_dustgasratio} that are absent from Fig.~\ref{fig:dg_fullcircle}, but these are simply due to spiral arms interacting with the periodic $\phi$-boundary conditions of a less-than-full annulus. The global picture with a gap, an inner and outer disc, streamers and a CPD remains the same.

\begin{figure}
\includegraphics[width=8.4cm]{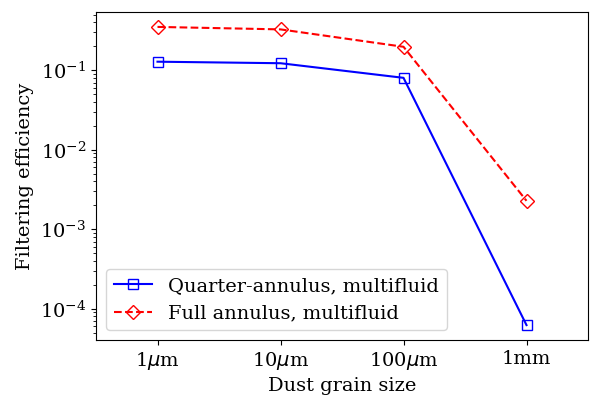}
\caption{Dust-to-gas mass ratio in the CPD normalised to the initial dust-to-gas mass ratio for that grain species. The blue, solid line is for the quarter-annulus multiple grain size simulation. The red, dashed line is for the full-annulus multiple grain size simulation.}
\label{fig:QCvsC_FE} \end{figure}

\begin{figure}
\includegraphics[width=8.4cm]{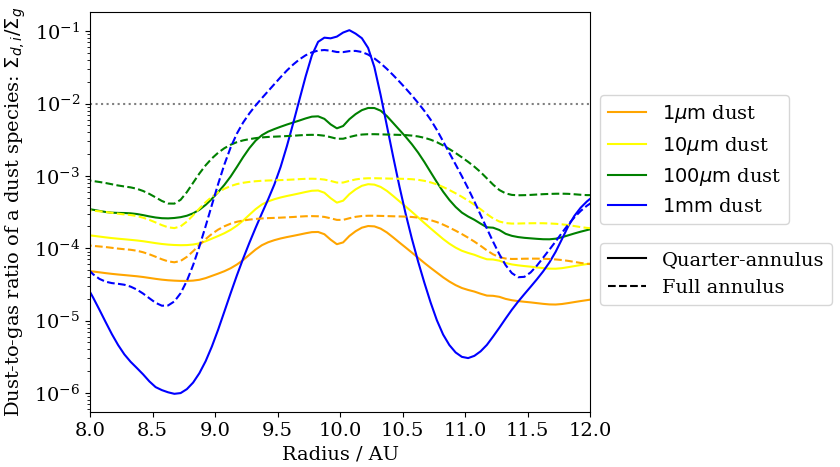}
\caption{Azimuthally and vertically averaged dust-to-gas mass ratio for the multiple grain size simulations, in steady state. That is at $t=50$ orbits for the quarter-annulus and $t=100$ for the full annulus.}
\label{fig:QCvsC_dgratiomulti} \end{figure}

Some quantitative difference can be observed between the quarter and full-annulus cases. From Fig.~\ref{fig:QCvsC_FE} it can be seen that the filtering efficiency for the different dust grain sizes, but especially $1\millim$, is much reduced (by a factor of 37, for $1\millim$) in the quarter-annulus case compared to the full-annulus case. This is because the protoplanet's gravitational torque is responsible for carving out the gap, by transferring orbital angular momentum from matter interior to its orbit to matter exterior of it. In effect, the quarter-annulus geometry exaggerates the time-integrated gravitational torque and the resulting planetary gap. Then, as more dust grains remain in the gap in the full-annulus case, more can be captured by the CPD. This effect of quarter-annulus geometry is stronger for larger grain sizes, a key weakness of simulations which depict less than the full annulus. However, note that size dependence of the filtering efficiency remains the same.

Also, the dust-to-gas mass ratio of the circumplanetary disc in the multiple grain size simulations -- considering dust of all grain sizes -- is $2.9 \times 10^{-4}$ for the quarter-annulus while $7.6 \times 10^{-4}$ for the full annulus. The diminution in the quarter-annulus case is likely due to the enhanced strength of gravitational torque discussed above. The torque particularly strongly affects $1\millim$ dust, which is the dominant dust-mass-carrier species. To visualise this effect whereby the quarter-annulus's enhanced torque exaggerates the gap compared to a full annulus, especially for large dust grains, see Fig.~\ref{fig:QCvsC_dgratiomulti}. Similarly, the ratio of the CPD dust mass to the protoplanet's mass is $7.0 \times 10^{-7}$ for the quarter-annulus while $4.5 \times 10^{-6}$ for the full annulus. Not only is the dust-to-gas ratio higher in the full annulus, but also the CPD gas mass (see Fig.~\ref{fig:QCvsC_FE}) because quarter-annulus geometry reduces the pool of available mass to accrete onto the CPD, which leads to this increase of a factor of 6.5.

Thus, while there are some quantitative difference between the quarter-and full annulus simulations, these differences are moderate and the overall behaviour and results remain the same.

\section{Discussion}
\label{sec:discussion}
\subsection{Benchmarking}
\label{sec:benchmarking}
Our simulations show a qualitatively similar picture as in the literature \citep[e.g.][]{KlahrKley2006,Machida2008,Tanigawa2012,Szulagyi2014}; i.e. the protoplanet carves a gap in the protoplanetary disc and, at the same time, a CPD forms around the protoplanet. The CPD structure itself shows a rotationally supported density structure filling the protoplanet's Roche lobe. Furthermore, the protoplanet accretes mass from the CPD while, at the same time, CPD material is replenished by meridional flows \citep{Szulagyi2014}. Previous studies do not, however, model the CPD with gas and multiple dust grain sizes each having their own dynamics.

\begin{figure}
\includegraphics[width=8.4cm]{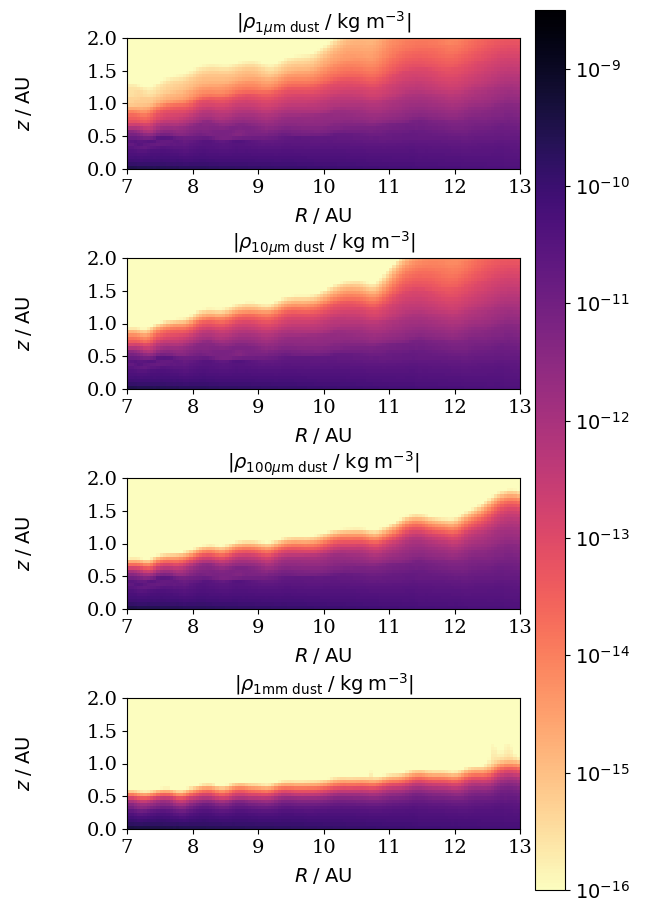}
\caption{Vertical slice at $\phi=0$ of the dust density, in units $\kg \metre^{-3}$, of a protoplanetary disc with no protoplanet yet inserted. Each subplot comes from a different single grain size simulation, with dust grain size $a=1\um$, $10\um$, $100\um$ and $1\millim$ from top to bottom.}
\label{fig:rhodust_singlegrain} \end{figure}

Dust grains also exhibit the expected behaviour: as seen in Fig.~\ref{fig:rhodust_singlegrain}, larger dust grains have a smaller vertical scale height \citep[e.g.][]{Nakagawa1986, Garaud2004, DullemondDominik2005, Fromang2006}. Small grains are strongly coupled to the gas by dust-gas drag and thus experience turbulent stirring, while large grains are weakly coupled and thus settle towards the midplane. Recently, observations have corroborated this picture. Observations of HL Tau show that $\sim \millim$ dust have a scale height of $H\sim 0.01R$ which is much flatter than for the gas disc. In contrast, \citet{Rich2021} find that, in the discs of IM Lup, HD 163296 and HD 97048, small-grain-size ($\sim \um$) dust at radii $< 100\AU$ has similar vertical distribution to the gas.

Furthermore, we also observe the effect of dust filtration, i.e. large grains are prevented from penetrating the annular gap, while small grains can flow in easily with the gas. This was already seen in 2D simulations \citep{Rice2006, Zhu2012,Weber2018, Haugbolle2019} which study the effect of dust clearing of the inner protoplanetary disc. They, however, do not consider the effect on the CPD.

\subsection{CPD grain size distribution}
\label{sec:grain_distribution}
Our results show that the grain size distribution function of the CPD follows the MRN distribution \citep{Mathis1977} at small grain sizes, but falls significantly below that distribution by grain size $a=100\um$ and is truncated to near zero by $a=1\millim$. Secondly, the total dust-to-gas ratio is significantly lower than the typical ISM value, i.e. $8\times 10^{-4}$ compared to $10^{-2}$. This is because of the lower dust content -- particularly of larger dust -- in the annular gap carved by the protoplanet due to dust filtration \citep{Zhu2012}. These results are similar to the findings of \citet{Bae2019} who used test particles for the dust grain dynamics in a 2D disc.

One caveat of this result is that the distribution function is only described by a limited number of size bins. It is necessary to increase the number of bins to examine the grain size distribution near the truncation. It is likely that there is no sharp transition, but a smooth turnover between $100\um$ and $1\millim$ as grains start to decouple from the gas. Another caveat is that, while we find that the dust distribution function follows an MRN power-law distribution at small grain sizes and tails off below MRN at $a=100\um$ to $1\millim$, this is only valid in absence of any grain processes. Local variations are expected as the grain size distribution is set by balancing dust production, growth and destruction processes. In protoplanetary discs, there is no dust production, but both growth and fragmentation of dust grains take place \citep{Brauer2008, Birnstiel2010, Birnstiel2018, Dullemond2018, Homma2018, Tamfal2018}. These restrictions are important when analysing observational data.

Understanding the grain size distribution function is also key to understanding observations as the distribution determines the opacity. When considering single-sized grains, \citet{Benisty2021} find that the CPD of PDS 70 c has a dust mass $0.007\Mearth$ if the grain size is $1\millim$ or $0.031\Mearth$ if $a=1\um$. Using an MRN distribution adjusted with the calculated filtering efficiencies (see Fig.~\ref{fig:filtering_efficiency}) via logarithmic linear interpolation to approximate the CPD's grain size distribution, we find that the CPD's opacity at wavelength $855\um$ -- and thus the CPD dust mass -- is close to that for single-sized grains of $1\millim$. However, as said before, with the coarse bin sizes there is some uncertainty on the turnover grain radius. Furthermore, \citet{Zhu2012} show that the critical grain radius for filtration depends on $\scr{\mathrm{St}}{crit}$ which is proportional to $\alpha$. It follows that $\scr{a}{crit}$ depends on the local temperature, viscosity and gas density of the disc: $\scr{a}{crit} \propto \scr{\nu}{turb} \rho_g / \left(c_{s,iso} \rho_m\right)$. We know that PDS 70 c lies further out in the protoplanetary disc at $34\AU$ and the local temperature is about $26\kelvin$ \citep{Benisty2021}, compared to $10\AU$ and $45\kelvin$ in our simulations. We need to assume some radial dependency for $\scr{\nu}{turb}$ and $\rho_g$. Let us assume that $\alpha$ is constant; then $\scr{\nu}{turb} \propto c_{s,iso}^2 R^{3/2}$. If we adopt $\rho_g \propto R^{-1}$, the critical grain size is 1.4 times that of our simulations, whereas if $\rho_g \propto R^{-3/2}$, $\scr{a}{crit}$ is 0.76 times ours. Either way, such a small variation in $\scr{a}{crit}$ makes little difference to the CPD dust mass deduced, as per Fig. 9 of \citet{Benisty2021}.

Via the method stated above, we can also calculate an opacity for our grain size distribution at $\lambda = 1.8\um$, the wavelength corresponding to the protoplanet's surface temperature $T=1600\kelvin$ by Wien's displacement law. It is $3.1$ times the opacity of the MRN distribution at the same wavelength. This can be understood by thinking of opacity as an absorption area-to-mass ratio. Our simulations include dust filtration to deplete the mass of large $\sim 1\millim$ grains, which have a lot of mass but little absorption area. Thus a CPD would be $>3$ times better at absorbing radiation emitted by its protoplanet than an MRN-distributed CPD and thus hotter, if it has the grain size distribution we obtain.

The reason why our results differ from those of \citet{SzulagyiBinkert2022}, who find that the CPD is enriched in dust compared to its parent PPD, is that they model the unperturbed PPD's dust as vertically flat, while our dust PPD is not flat because we do not neglect turbulent diffusion as they do. They find that large dust grains can accrete efficiently onto the protoplanet because the protoplanet vertically stirs up their flat disc of dust, pushing dust to high altitude where it can flow to feed the protoplanet, when flows at the midplane cannot do so because large dust grains are blocked as per dust filtration \citep[e.g.][]{Haugbolle2019}. We find, contrarily, that the protoplanet \textit{pulls down} the dust towards the midplane. An alternative reason is that their $\scr{a}{crit}$ is larger than ours, so large dust grains are still small enough not to be blocked off. However, their $\scr{a}{crit}$ is only larger than ours for their $5.2\AU$ case, not their $30\AU$ and $50\AU$ cases.

\subsection{Dust mass and satellite formation}
\label{sec:grain_growth}
With the \citet{Benisty2021} $0.007\Mearth$ estimate of the CPD dust mass for PDS 70 c, the ratio of the CPD dust mass to protoplanet mass is about $10^{-5}$ where we assume the protoplanet mass to be $2\Mjup$ \citep{Wang2020}. In our simulation -- meaning the full-annulus multiple grain size simulation, the most physically realistic one -- the ratio is even lower than that: $4.5 \times 10^{-6}$. This discrepancy is simply the result of differing temperature assumptions. For the same observed flux, the higher the assumed temperature, the lower the deduced mass. \citet{Benisty2021} assume a CPD temperature of $26\kelvin$, whereas the mass-averaged temperature of the CPD in our full-annulus multiple grain size simulation is $105\kelvin$. This is likelier to be an underestimate of the temperature than an overestimate, because we may include the protoplanet's luminosity but we neglect shock heating from the matter falling vertically at up to $15\km \ \mathrm{s}^{-1}$ towards the CPD. Furthermore, \citet{Isella2019} observe a similar flux for PDS 70 c's CPD to \citet{Benisty2021} and they estimate its dust mass at $0.004\Mearth$ if $T \sim 20\kelvin$ and $0.002\Mearth$ if $T \sim 80\kelvin$. With a temperature in the latter case more similar to ours, they obtain a CPD dust to protoplanet mass ratio of $3.2 \times 10^{-6}$. This brings our simulated value of the mass ratio and the value inferred from observed flux within decent agreement.

If our numerical model is overestimating the accretion rate from the CPD onto the protoplanet, the true mass of the CPD may be greater than we calculate. Our simulation result should be understood as providing a lower limit for CPD dust mass, rather than exact. This observational comparison provides support that it is at least reasonable on an order-of-magnitude basis.

The major satellites of Jupiter, combined, have a mass $\sim 2 \times 10^{-4}$ times the mass of their host planet, and the same ratio holds true for Saturn \citep{CanupWard2009}. However, the CPD dust mass is determined by the balance of removal through accretion of the protoplanet and replenishment from the protoplanetary disc. This is different than for a protoplanetary disc as the CPD is embedded in a gas and dust reservoir, while the protoplanetary disc is not and can become depleted of dust \citep{CanupWard2002}. Because of this, it is not actually necessary for the instantaneous dust mass of the CPD at any one moment to be high, for satellites to be formed. This is known in the literature as the `starved disc' model as discussed in Sect.~\ref{sec:introduction}. Also, planetesimal capture can provide satellitesimal seeds for this dust to accrete onto \citep[e.g.][]{Ronnet2020} and \citet{Drazkowska2018} show that dust traps are an efficient way to form satellites within the CPD. This would limit the accretion of dust onto the protoplanet and make more dust available to accumulate and form satellites. However, our simulations do not have enough resolution to follow the detailed evolution of the CPD and its satellite formation process. It thus is beyond the scope of this paper to follow the detailed evolution of the CPD involving the formation of satellites.

Furthermore, both the CPD dust to protoplanet mass ratio and the CPD dust-to-gas mass ratio can be expected to be higher at earlier times in planet formation. For these simulations, recall that the mass of the parent protoplanetary disc was set to $0.05\Msun$ around a $1\Msun$ star. If gas density is higher, there is a larger $\scr{a}{crit}$ (Sect.~\ref{sec:grain_distribution}). Most of the dust mass is in larger grains, so when the critical grain size is larger, much more of the dust mass is able to enter the CPD in spite of dust filtration. And of course, for a protoplanet which has not (or not yet) grown massive enough to carve out a gap in the PPD, the CPD dust-to-gas mass ratio can be very much higher than we find here, as the gap's dust filtration effect is absent.

\section{Conclusions}
\label{sec:conclusions}
We run 3D hydrodynamical simulations of a segment of protoplanetary disc with an embedded Jupiter-mass protoplanet orbiting a Solar-mass star at orbital radius $10\AU$. We follow the dynamics of the gas and 4 different dust grain sizes ($1\um$, $10\um$, $100\um$ and $1\millim$). We include the effects of turbulent viscosity and dust-gas drag, using either the Epstein or the Stokes drag law depending on the ratio of the dust grain size to the gas's mean free path. We include the back-reaction due to the drag force of the dust on the gas. The different dust grain sizes are not coupled directly by a force, but via their back-reaction on the gas, they can indirectly influence each other. This is the first time multiple dust grain sizes with separate dynamics have been simulated in a CPD.

We obtain the following conclusions:
\begin{enumerate}
\item  The dynamics of the grains in the multiple grain size simulation is similar to the dynamics observed in the single grain size simulations. As the large grains modify the gas dynamics due to the back-reaction of dust-gas drag, they also modify the dynamics of the small grains. However, these changes are not significant and do not affect the CPD.
\item At small grain sizes $< 100\um$, the grain size distribution of the dust in the CPD shows an MRN distribution. It tails off significantly below MRN at $a=100\um$ and falls to almost zero by $a=1\millim$, due to dust filtration limiting the flow of large dust grains into the annular gap. The critical grain radius for dust filtration depends on the local properties of the disc, i.e. the disc density, temperature and viscosity.
\item The CPD is depleted in dust-to-gas ratio compared to the parent protoplanetary disc by an order of magnitude, but is similar to the value within the annular gap carved by the protoplanet.
\item Because the truncation and the low dust-to-gas ratio in the CPD, the CPD dust mass is low. The ratio of the CPD dust mass to the protoplanetary mass is $\sim$ a few $\times 10^{-6}$. While this is considerably lower than the value of  $2\times 10^{-4}$ of Jupiter's mass that constitutes the total mass of its moons, the dust within the CPD is continuously replenished by dust flow from the protoplanetary disc, thus making satellite formation possible as per the `starved disc' model in the literature.
\item The opacity, mass-averaged temperature, and CPD dust to protoplanet mass ratio derived from our multiple grain size simulation yield consistency with the fluxes observed from the CPD of PDS 70 c by \citet{Isella2019} and \citet{Benisty2021}.
\end{enumerate}
Our simulations consider only a singular environment while changing the dust distribution between simulations. To further understand how environmental conditions change the grain size distribution, we need to change these parameters. In a subsequent study we will consider different planetary masses and position within the protoplanetary disc and also consider finer size binning in order to refine the critical grain size affected by dust filtration.

\section*{Acknowledgements}
S.M.K. acknowledges funding from the Royal Society through the Fellowship Enhancement Award (grant holder O.P.). The research of O.P. was supported by the Royal Society Dorothy Hodgkin Fellowship during the preparation of this publication. S.v.L. is supported by a STFC consolidated grant. S.M.K. would like to thank Dr James Miley for the protoplanetary disc models that underlie these simulations. This work was undertaken on ARC4, part of the High Performance Computing facilities at the University of Leeds, UK.

\section*{Data Availability}
The data underlying this article will be shared on reasonable request to the corresponding author.

\bibliographystyle{mnras}
\bibliography{paper1} % for a bibtex file called paper1.bib

\appendix
\section{Accretion algorithm}
\label{sec:accretion}
We wrote a Gaussian accretion algorithm, designed to prevent sharp, discontinuous, un-physical transitions for a protoplanet moving across a grid of cells. The method bears some resemblance to, but is not identical to, that of \citet{Krumholz2004}. The amount of matter accreted from a cell containing density $\rho$ in each timestep $\Delta t$ is given by:
\begin{equation}
\Delta m = f \rho \scr{V}{cell} \times \left( 1 - \func{\mathrm{exp}}{\frac{- \Delta t}{ \scr{t}{acc} }} \right)
\func{ \mathrm{exp} }{ \frac{- \left| \ve{r} - \scr{\ve{r}}{pl}\right|^2 }{ r_G^2 } }
\label{eq:accretion} \end{equation}
where $f$ is an order-unity constant and $r_G$ is the `Gaussian radius' of the protoplanet, which is chosen to be $r_G = 3 \scr{R}{eff}$. Here $\scr{R}{eff}$ is the effective radius that was defined in Sect.~\ref{sec:temperature}. As $\Delta t \ll \scr{t}{acc}$ in practice, the amount of mass accreted from a cell in time $\Delta t$ is proportional to $\Delta t$. This is deliberate; a conclusion for the accretion rate should not depend on the user's arbitrary numerical timestep. The accretion timescale works like a freefall timescale: $\scr{t}{acc}^2 = \pi^2 R_{ff}^3 / \left(8G\Mpl\right)$, where $R_{ff} = \func{\mathrm{max}}{\left| \ve{r} - \scr{\ve{r}}{pl}\right|, \scr{R}{eff}}$. The truncation of distance from the protoplanet at minimum value $\scr{R}{eff}$ is to avoid a singularity at the position of the protoplanet.

This accretion is applied separately to the gas and to every species of dust. Whenever the protoplanet accretes matter from a cell, it records -- separately -- how much gas and how much dust it has accreted. This enables the simulations to track how efficiently the protoplanet accretes dust, by comparison to its accretion of gas.

Following \citet{Krumholz2004}, it is not advisable to let the sink particle violate the conservation of angular momentum around it when it accretes matter onto itself. Accordingly, whenever accretion is carried out for a fluid in the cell, the velocity of that fluid in that cell is decomposed into a component comoving with the protoplanet and the remainder `peculiar' velocity. The peculiar velocity vector is further decomposed using a cell-specific spherical coordinate system centred on the protoplanet, with unit-vectors $\left(\unitvector{r,\mathrm{pl}}, \unitvector{\phi,\mathrm{pl}}, \unitvector{\theta,\mathrm{pl}}\right)$. Hence $\ve{v} = \scr{\ve{v}}{pl} + \Sigma_i v_{\mathrm{rel},i}$ where we define $v_{\mathrm{rel},i} = \unitvector{i,\mathrm{pl}} \ \ve{.} \left( \ve{v} - \scr{\ve{v}}{pl} \right)$ where $i \in \{r,\theta,\phi\}$. When some mass is removed from the cell onto the sink particle, the component of momentum comoving with the protoplanet $m\scr{\ve{v}}{pl}$ and the peculiar component $mv_{\mathrm{rel},r}$ are accreted, whereas the peculiar components $mv_{\mathrm{rel},\theta}$ and $mv_{\mathrm{rel},\phi}$ are conserved during accretion. If mass of a fluid $\Delta m$ is accreted from a cell, the momentum of that same fluid accreted from the same cell is $\Delta \ve{p} = \left( \scr{\ve{v}}{pl} + v_{\mathrm{rel},r} \unitvector{r,\mathrm{pl}} \right) \times \Delta m$.

% Don't change these lines
\bsp	% typesetting comment
\label{lastpage}
\end{document}